\pdfoutput=1

\documentclass{JINST}

\title{Development of a Li$_2$MoO$_4$ scintillating bolometer for low background physics}

\author{
L.~Cardani$^{c,d}$,
N.~Casali$^{a,b}$,
S.~Nagorny$^a$,
L.~Pattavina$^a$\thanks{Corresponding
author.}~, 
G.~Piperno$^{c,d}$
O.P.~Barinova$^e$, J.W.~Beeman$^f$, F.~Bellini$^{c,d}$, F.A.~Danevich$^g$, S.~Di~Domizio$^{h,i}$, L.~Gironi$^{j,k}$, S.V.~Kirsanova$^e$, F.~Orio$^d$, G.~Pessina$^{j,k}$, S.~Pirro$^a$, C.~Rusconi$^k$, C.~Tomei$^d$, V.I.~Tretyak$^g$ and M.~Vignati$^d$\\
\llap{$^a$}INFN - Laboratori Nazionali del Gran Sasso, I-67010 Assergi (AQ) - Italy\\
\llap{$^b$}Dipartimento di Scienze Fisiche e Chimiche - Universit\`{a} degli studi dell'Aquila, I-67100 Coppito (AQ) - Italy\\
\llap{$^c$}Dipartimento di Fisica - Universit\`{a} di Roma La Sapienza, I-00185 Roma - Italy\\
\llap{$^d$}INFN - Sezione di Roma, I-00185 Roma - Italy\\
\llap{$^e$}D.I. Mendeleev University of Chemical Technology of Russia, 125047 Moscow - Russia\\
\llap{$^f$}Lawrence Berkeley National Laboratory , Berkeley, California 94720 - USA\\
\llap{$^g$}Institute for Nuclear Research, MSP 3680 Kyiv - Ukraine\\
\llap{$^h$}Dipartimento di Fisica, Universit\`{a} di Genova, I-16146 Genova - Italy\\
\llap{$^i$}INFN - Sezione di Genova, I-16146 Genova - Italy\\
\llap{$^j$}Dipartimento di Fisica - Universit\`{a} di Milano Bicocca, I-20126 Milano - Italy\\
\llap{$^k$}INFN - Sezione di Milano Bicocca, I-20126 Milano - Italy\\

  E-mail: \email{luca.pattavina@lngs.infn.it}

  }

\abstract{We present the performance of a 33~g Li$_2$MoO$_4$ crystal working as a scintillating bolometer. The crystal was tested for more than 400~h in a dilution refrigerator installed in the underground laboratory of Laboratori Nazionali del Gran Sasso (Italy). This compound shows promising features in the frame of neutron detection, dark matter search (solar axions) and neutrinoless double-beta decay physics. Low temperature scintillating properties were investigated by means of different $\alpha$, $\beta$/$\gamma$ and neutron sources, and for the first time the Light Yield for different types of interacting particle is estimated. The detector shows great ability of tagging fast neutron interactions and high intrinsic radiopurity levels (<~90~$\mu$Bq/kg for $^{238}$U and <~110~$\mu$Bq/kg for $^{232}$Th).}


\begin{document}

\section{Introduction}

Lithium molybdate (Li$_2$MoO$_4$) crystal scintillators can have important applications in applied physics
(detection of neutron fluxes through $n$ captures by $^6$Li) as well as in fundamental physics in searches
for neutrinoless double-beta decay (DBD), dark matter (DM) and solar $^7$Li axions.

Double-beta decay is a process of rare nuclear transformation $(A,Z)$ $\to$ $(A,Z\pm2)$ with simultaneous
emission of two electrons or positrons. Two neutrinos ($2\nu$) DBD, when two neutrinos are also emitted,
is allowed in the Standard Model (SM) of particle physics but, being the second order process in weak
interactions, this is the rarest nuclear decay observed to-date, with half-lives in the range of
$T_{1/2}$ $\simeq$ $10^{18}-10^{24}$ yr \cite{Ver12,Giu12,Gom12}. Neutrinoless ($0\nu$) DBD is forbidden in
SM because it violates the lepton number conservation; however, it is predicted by many SM extensions
which describes neutrino as Majorana particle ($\nu=\overline{\nu}$) with non-zero mass.
$^{100}$Mo is one of the favorite isotopes in searches for $0\nu$DBD because of high energy release
$Q_{\beta\beta}=3.034$ MeV and quite big natural abundance $\delta=9.82\%$ which can be relatively
cheaply increased by centrifugal method. Only $T_{1/2}$ limits for $0\nu$DBD are known to-date,
with the best value for $^{100}$Mo reached in the NEMO-3 experiment $T_{1/2}>1.0\times10^{24}$ yr
\cite{Sim12}. Scintillating bolometers are considered now as one of the most perspective tools in
high-sensitive searches for $0\nu$DBD because of high efficiency (realized in the ``source = detector''
approach) and possibility to distinguish $\beta\beta$ signal from backgrounds ($\alpha$ decays, pile-ups,
etc.) by simultaneous measurement of the heat and light channels \cite{Pir06}.
Recently, several compounds containing Mo have been studied~\cite{Lee11}~\cite{ZnMoO4}~\cite{PbMoO4}, all of them have shown a great ability to identify $\alpha$ interactions from $\beta$/$\gamma$ ones. This feature is extremely important because, as demonstrated by the CUORICINO collaboration~\cite{QINO}, in DBD bolometric experiments, one of the primary background sources are degraded $\alpha$ particles interacting with the detector. By means of scintillating properties these types of interactions can be easily identified and discarded.
High-scale projects of $^{100}$Mo $0\nu$DBD search with sensitivity $T_{1/2}$ $\simeq$ $10^{26}$ yr
are known with two Mo-containing scintillating bolometers: CaMoO$_4$ \cite{Lee11} and ZnMoO$_4$
\cite{ZnMoO4,Bee12b}. Appearance of new highly radiopure scintillator in this field can be very
interesting. Achievements in technology allow to grow now by Czochralski technique quite large
($\oslash 3\times6$ cm) transparent Li$_2$MoO$_4$ crystals \cite{Bar13}, and further progress is expected.\newline

In accordance with our contemporary understanding of astronomical observations, usual matter constitutes
only $\simeq4\%$ of the Universe, while the main components are dark matter ($\simeq23\%$) and dark energy
($\simeq73\%$) \cite{DM_review,Fen10}. Quite big number of candidates were proposed on the role of
weakly interacting massive particles (WIMPs) from which the DM could be composed; one of the approaches
to discover these particles is to detect their scattering on nuclei in sensitive detectors placed deep
underground in low background conditions. Highly radiopure scintillating bolometers with low energy
threshold which have an ability to distinguish nuclear recoils from other signals ($\beta/\gamma$, $\alpha$,
noises, etc.) are very perspective tools in this field too.\newline

Li$_2$MoO$_4$ can be used also in searches for $^7$Li solar axions. Axions are consequence of the
Peccei-Quinn solution of the so-called strong CP problem in quantum chromodynamics \cite{Kim10}.
If axions exist, the Sun can be an intensive source of these particles. In particular, they could be
emitted instead of $\gamma$ quanta in M1 transitions which depopulate the excited level of $^7$Li with
$E_{exc}=477.6$~keV; this level is populated in the main $pp$ cycle of nuclear reactions in the solar
interior. Coming to the Earth, such axions could resonantly excite the same level of $^7$Li in some
Li-containing target \cite{Resonant_axions,Li7_axions}. Up to now, Ge detectors and external Li targets were used to
search for de-excitation $\gamma$ quanta with energy of 477.6 keV (see \cite{Bel12} and Refs. therein);
this restricted efficiency of detection to a few \%. Use of the ``source = detector'' approach with
big Li$_2$MoO$_4$ would increase efficiency and thus sensitivity to these hypothetical particles.

Radiopurity of Li$_2$MoO$_4$ crystal, as external source in
measurements with HPGe detector, was studied in \cite{Radio_LiMo}; it
was found quite radiopure. In \cite{First_LiMo}, a small $\oslash25
\times 0.9$~mm Li$_2$MoO$_4$ crystal was tested at the first time
as a cryogenic scintillating bolometer, and both light and heat
signals were observed; this demonstrated that Li$_2$MoO$_4$ can be
an interesting candidate in low background physics. Here we
present results of further studies of a larger $\oslash22 \times
33$~mm Li$_2$MoO$_4$ crystal sample as scintillating bolometer in
underground measurements in the Laboratori Nazionali del Gran
Sasso (3600 m w.e.) with different $\alpha$, $\beta/\gamma$ and
neutron sources.

\section{Experimental set-up}

A Li$_2$MoO$_4$ cylindrical crystal was operated for more than 400~h together with a Ge Light Detector (LD) as a scintillating bolometer, inside a dilution refrigerator in the underground laboratory of Laboratori Nazionali del Gran Sasso (Italy).\newline

\begin{figure}[t]
\centering
\includegraphics[width=0.5\textwidth]{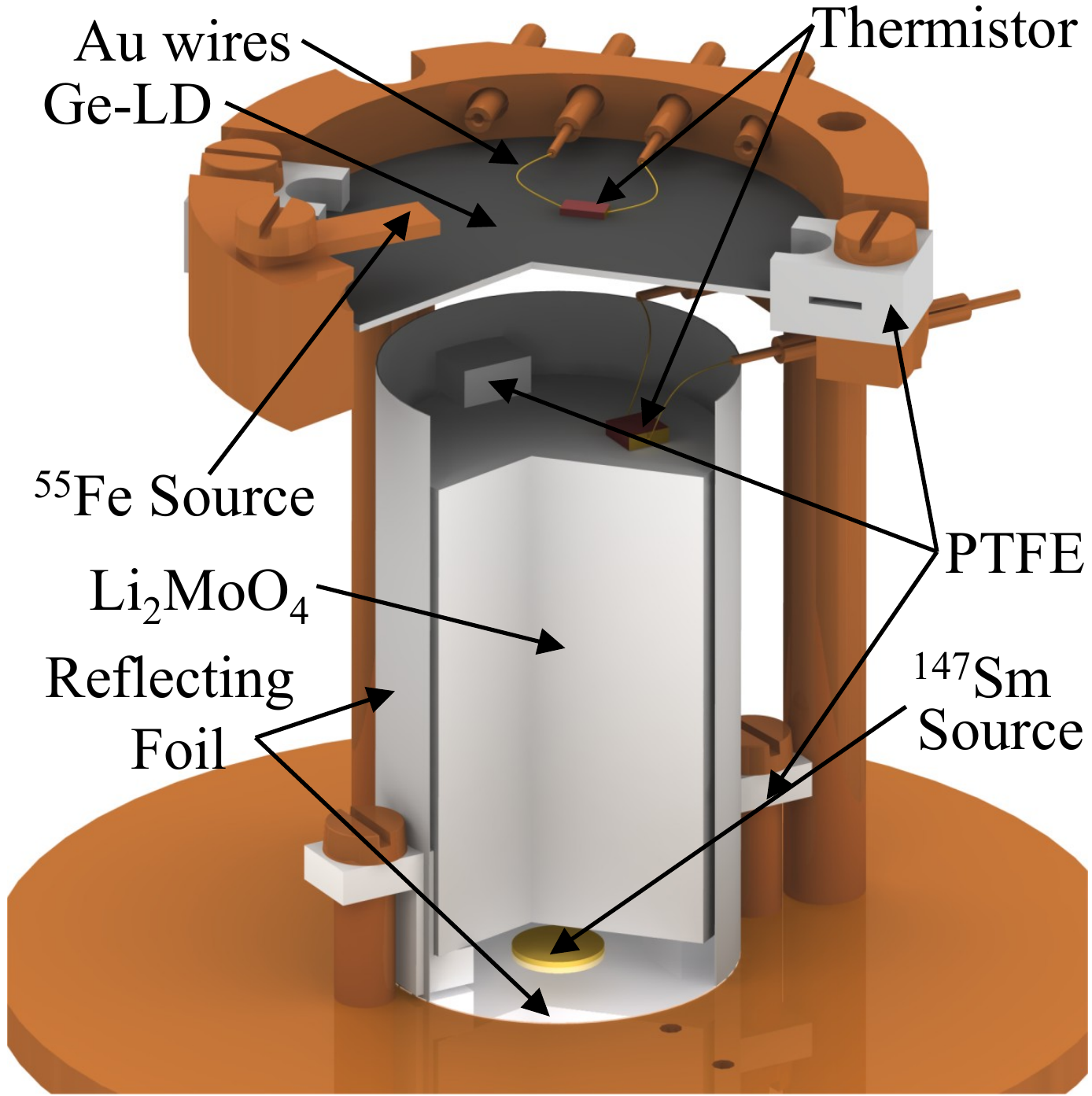}
\caption{Set-up of the detectors. The Li$_2$MoO$_4$ is held in position by PTFE $S$-shaped supports. The Ge Light Detector is facing the main crystal in such a way to maximize the light collection. The electrical signals are read from the ground-isolated Cu pins that are connected to the thermistors by means of Au wires.}
\label{fig1} 
\end{figure}

The crystal (\o=22~mm and h=33~mm) is held in position by means of four PTFE supports fixed on four copper columns, two are pulling up the crystal from the bottom surface and two are pushing it down from the top face. In order to maximize the light collection on the LD, the structure is surrounded by a reflecting foil (3M VM2002), employed as a light guide. The LD consists of a hyper-pure Ge disk 36~mm of diameter and 1~mm thickness, it is coupled to a copper structure by means of a pair of PTFE clamps (see Fig.~\ref{fig1}). The Ge detector is also operated as a bolometer, more information about operational characteristics of bolometric LD can be found in~\cite{LD}.\newline 

The two absorbers are equipped each one with a temperature sensor: a Neutron Transmutation Doped (NTD) germanium thermistor, similar to the one used in Cuoricino~\cite{QINO}, that converts the thermal signal into an electrical pulse. The thermometers are thermally coupled to the absorbers by means of epoxy glue spots (\o~$\sim$~0.5~mm and h~$\sim$~50~$\mu$m). On the Li$_2$MoO$_4$ crystal the sensor dimensions are 3$\times$3$\times$1~mm$^3$, while on the LD 3$\times$1.5$\times$0.1~mm$^3$.

\section{Detector operation and data analysis}

The thermistors of the detectors are biased with a constant current, the resistance variations, generated by the temperature rise, are converted into voltage signals by the sensors. The heat and light signals are then amplified, filtered by a 6-pole Bessel-filter with a cut-off  frequency at 120~Hz, and at the end fed into a NI PXI-6284 18-bits ADC.\newline

Triggers are software generated, they ensure that each pulse is recorded with a 1~kHz sampling rate. When the triggers fire, for the main bolometer and the LD, waveforms 1~s and 0.25~s long are recorded on disk. Moreover, when the trigger of the Li$_2$MoO$_4$ crystal fires, the corresponding waveform from the LD is recorded, irrespective of its trigger. The amplitude and the shape of the voltage pulse is then determined by the off-line analysis. The pulse amplitude of the thermal signals are estimated by means of the Optimum Filter (OF) technique~\cite{Gatti}\cite{Radeka}, that maximizes the signal-to-noise ratio in a way to increase the energy resolution of the detector. The amplitude of the scintillation light signals, however, is evaluated from the filtered waveforms at a fixed time delay with respect to the Li$_2$MoO$_4$ bolometer, as described in detail in~\cite{light_sync}.\newline
The amplitude of the heat pulses is energy-calibrated using different known peaks in the acquired spectrum.\newline

The detector was operated for more than 400~h, during the data taking calibration sources were used in order for studying the detector response to different types of interactions and to estimate the Light Yield (LY) for different particles energy deposit. $^{137}$Cs and $^{40}$K were employed as $\gamma$ sources, placing those in proximity of the experimental set-up. Natural Sm was deposited on the surface of the reflecting foil facing the Li$_2$MoO$_4$ crystal on the opposite side of the LD, $^{147}$Sm was used as $\alpha$ source for illuminating the crystal. Finally an Am-Be neutron source was placed close to the detector for studying the neutron capture reaction on $^6$Li.\newline

The LD performances were evaluated by means of a $^{55}$Fe X-ray source close to the detector so that to illuminate homogeneously the Ge face. The two lines at 5.9~keV and at 6.5~keV, produced by the source, were used for the computation of the FWHM of the LD and for a calibration of the LY of the Li$_2$MoO$_4$ crystal. The detector, the same used in~\cite{PbWO4}, showed good performance in terms of energy resolution. The FWHM energy resolution at the $^{55}$Fe X-rays was 262~eV.

%

\subsection{Calibration with $\gamma$ sources}
\label{gamma_calib}
The crystal was exposed to $^{137}$Cs and $^{40}$K $\gamma$ sources in order to evaluate the energy resolution of the thermal channel of the detector and also to estimate the LY for $\beta$/$\gamma$ interactions. We define the LY as the ratio between the measured scintillation light (keV) produced by a $\beta$/$\gamma$ event in the crystal and the nominal energy of that event (MeV). The LY for $\beta$/$\gamma$ events was evaluated averaging the LY of $^{137}$Cs and $^{40}$K lines, the computed value is:
\begin{equation}
LY_{\beta / \gamma} = 0.433 \pm 0.012 \; keV/MeV.
\end{equation}
This is the first time that the LY$_{\beta / \gamma}$ for this compound is evaluated at low temperature and it is in agreement with the estimation of~\cite{First_LiMo}.\newline

\begin{figure}[t]
\centering
\includegraphics[width=0.75\textwidth]{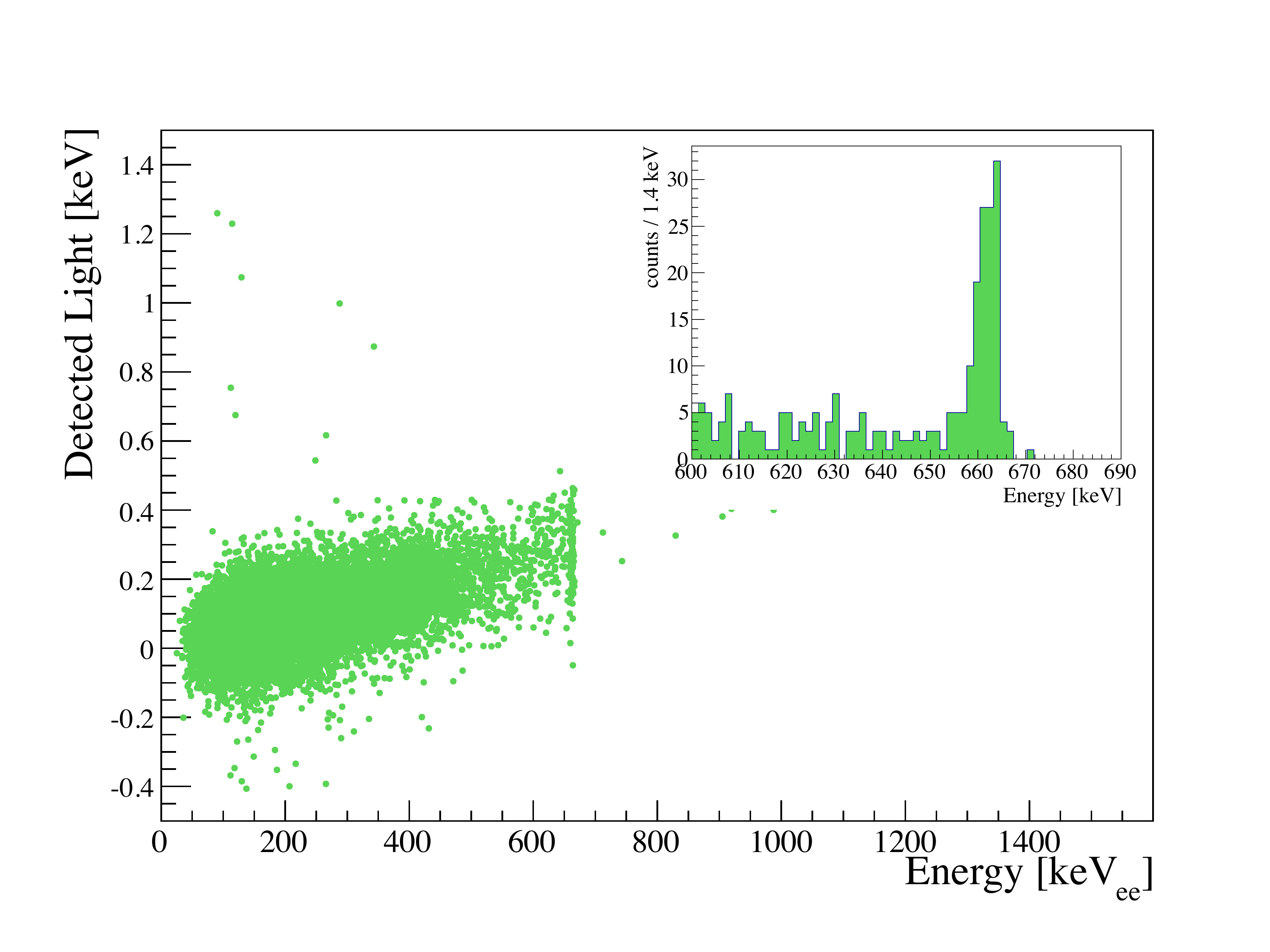}
\caption{Light vs. Heat scatter plots for a calibration measurement with a $^{137}$Cs source. In the inset the source peak is shown, where just $\beta$/$\gamma$ events are selected from the Heat channel.}
\label{fig:cs137} 
\end{figure}

\begin{figure}[t]
\centering
\includegraphics[width=0.75\textwidth]{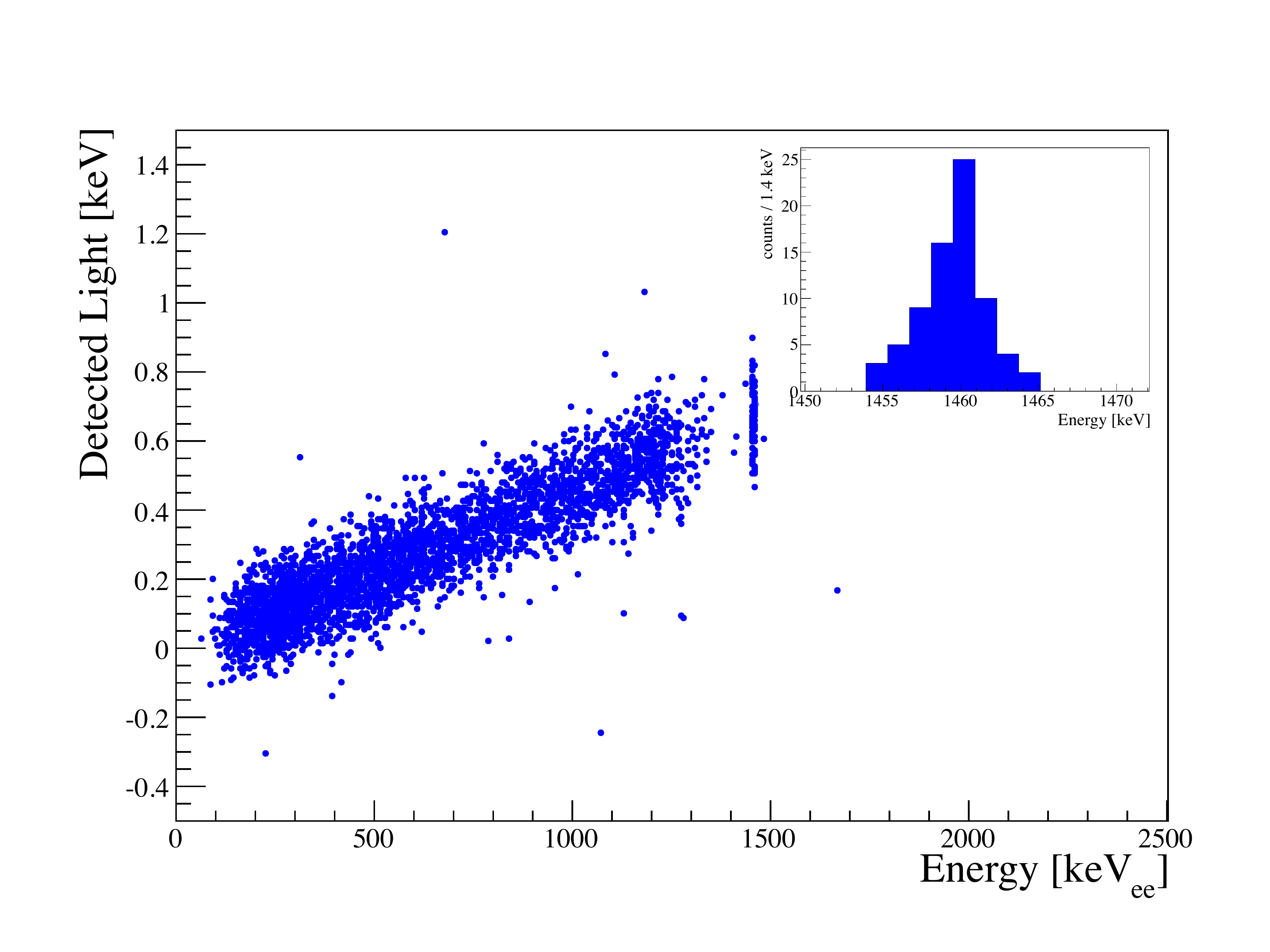}
\caption{Light vs. Heat scatter plots for a calibration measurement with a $^{40}$K source. In the inset the source peak  is shown, where just $\beta$/$\gamma$ events are selected from the Heat channel.}
\label{fig:k40} 
\end{figure}

The FWHM energy resolution of the Li$_2$MoO$_4$ crystal is evaluated and it varies from 3.9$\pm$0.8~keV at 661~keV to 4.7$\pm$1.1~keV at 1460~keV (see Fig.~\ref{fig:cs137} and Fig.~\ref{fig:k40}).

\subsection{Calibration with a neutron source}
We decided to expose our set-up to an intense neutron source, about 200~$n$/s. An Am-Be neutron source was placed next to our experimental set-up. In fact the compound is extremely interesting for neutron spectroscopy~\cite{Lithium-6}, due to the relatively high neutron absorption cross section of $^6$Li, both for thermal (940~b at 25~meV) and fast (10~b at 240~keV) neutrons.\newline

The aim of the measurement was to study the reaction:
\begin{equation}
\label{reaction}
^6Li \; + \; n \; \rightarrow \; ^3H \; + \; ^4He \; +\; 4.78~MeV.
\end{equation}
After the absorption of a neutron by a $^6$Li nucleus there is the emission of $^3$H and $^4$He nuclei. The kinetic energy of the reaction products is 4.78~MeV plus the kinetic energy of the neutrons. $^6$Li is an interesting nuclide because by means of its high absorption cross section and reaction allows neutron spectroscopy, even for high energy neutrons.\newline

\begin{figure}[h]
\centering
\includegraphics[angle=0, width=0.75\textwidth]{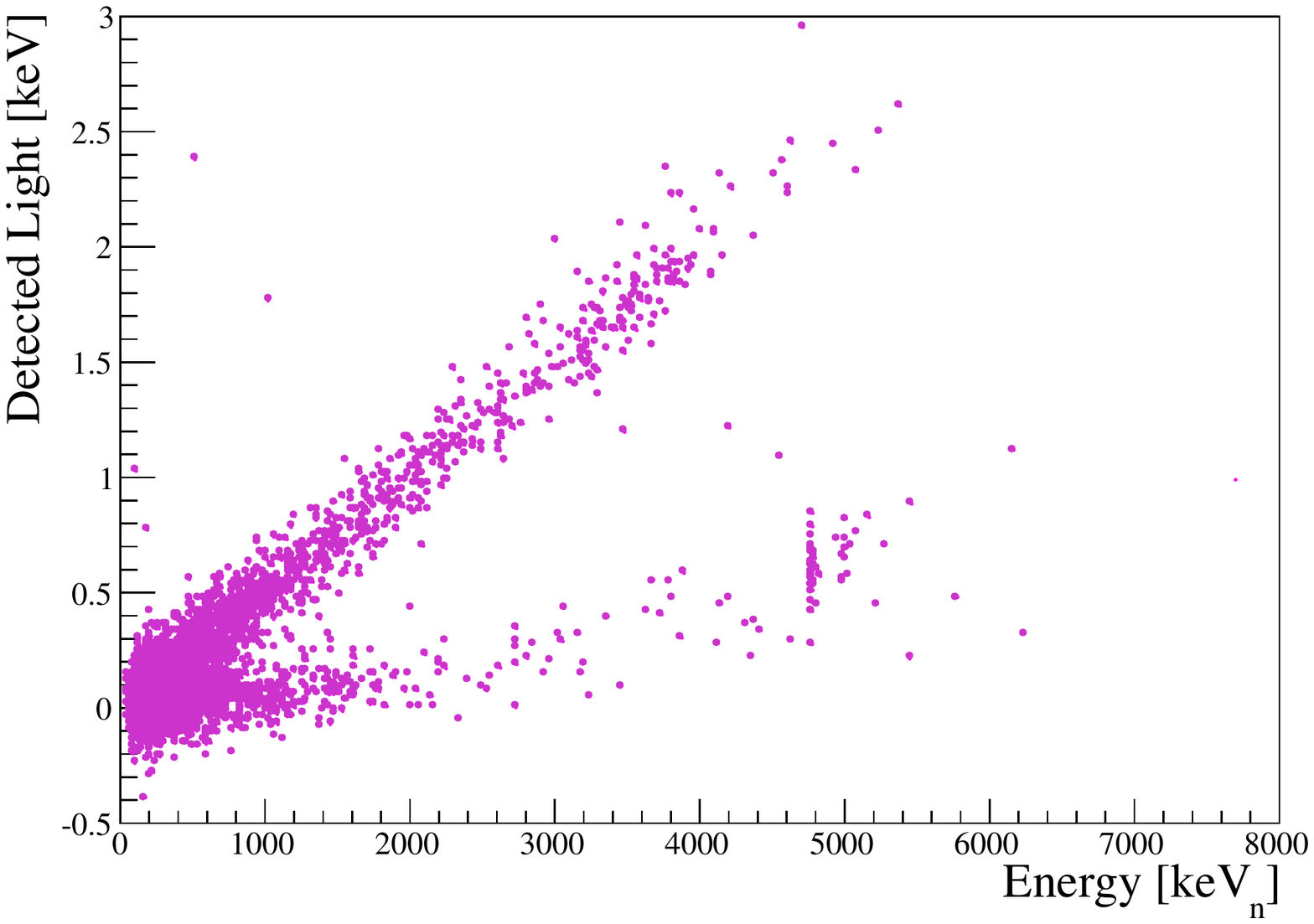}
\caption{Light vs. Heat scatter plot for a calibration measurement with a Am-Be neutron source. The larger LY band is ascribed to $\beta$/$\gamma$ events produced by the interactions of neutrons with the materials surrounding the detector, while in the lower one there are direct neutron interactions in the absorber.}
\label{fig:n} 
\end{figure}

\begin{figure}[h]
\centering
\includegraphics[angle=0, width=0.75\textwidth]{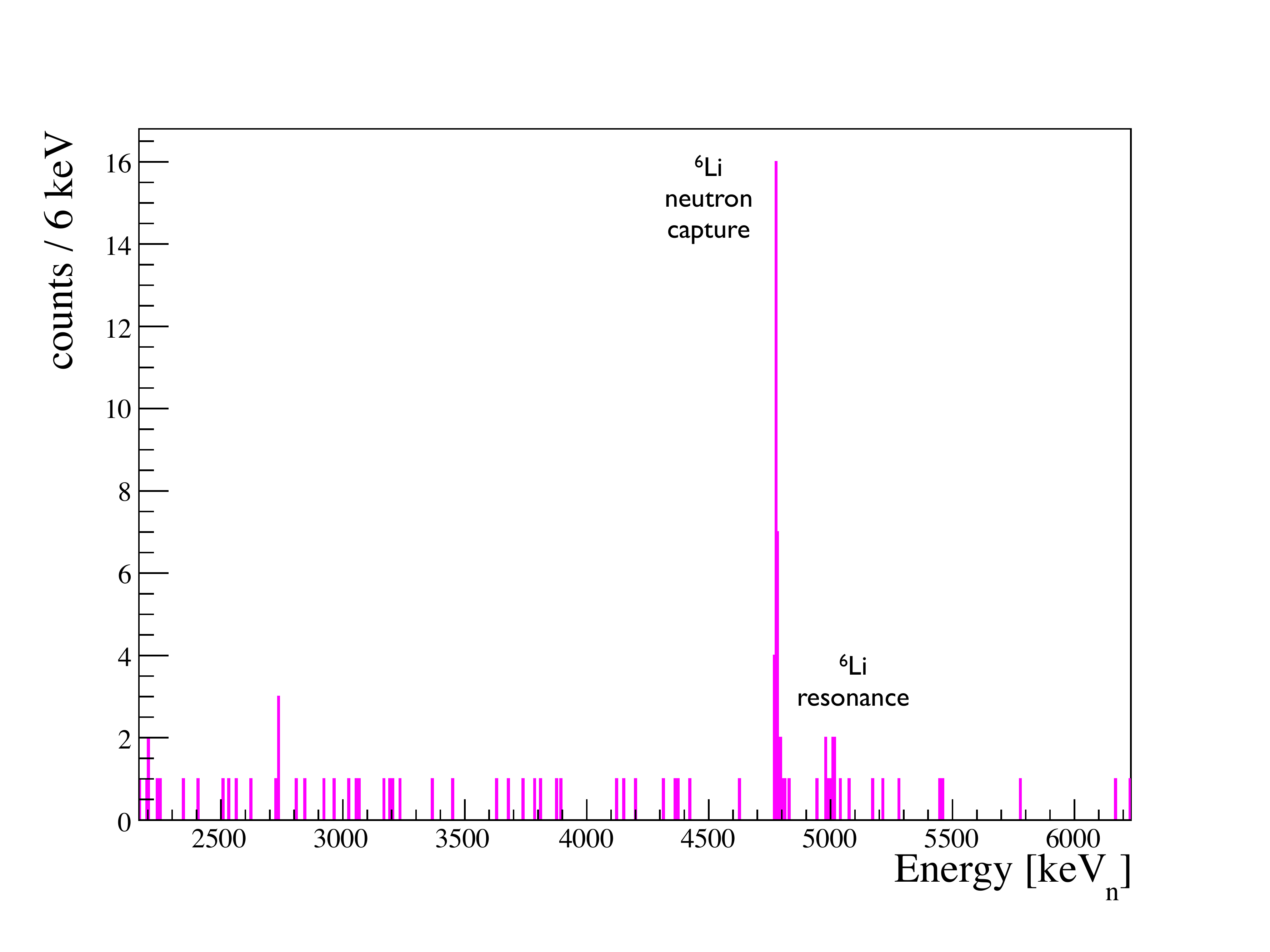}
\caption{Energy spectrum of low light yield events (lower band of the Light vs. Heat scatter plot for the calibration neutron measurement). The most intense peak at 4.78~MeV is ascribed to the $^6$Li(n$_{thermal}$,$\alpha$)$^3$H reaction, while the cluster of events at 5~MeV is produced by the resonance neutron capture still on $^6$Li. }
\label{fig:h_n} 
\end{figure}

The live time of the calibration measurement was 18~h, in Fig.~\ref{fig:n} is shown the Light vs. Heat energy scatter plot.
The graph is characterized by a peculiar feature, a line at 4.78~MeV, produced by the energy deposit of $^3$H and $^4$He nuclei in the crystal, these are reaction products of $^6$Li thermal neutrons absorption. If we apply a cut on the Light channel selecting just events in the low LY band (thus excluding $\beta$/$\gamma$ ones), we obtain the energy spectrum of Fig.~\ref{fig:h_n}, which consists of just neutron energy deposits in the absorber. The energy spectrum shows various signatures produced by neutron interactions:
\begin{itemize}
\item nuclear recoils produced by fast neutrons scattering on the different nuclei of the compound, which are represented by the continuum of events that extends up to 4.8~MeV. Large energy deposits are produced by the scattering on light nuclei, e.g. on Lithium and Oxygen isotopes, where neutrons can lose a large fraction of their kinetic energy;
\item thermal neutrons absorption peak at 4.78~MeV;
\item fast neutron absorptions on $^6$Li nuclei, which are represented by events with energy greater than 4.78~MeV. The absorption cross section shows a resonance at about 240~keV~\cite{XS_Li6}, so the excess of events at around 5.02~MeV can be ascribed to this type of interactions ($Q$-value of the reaction + kinetic energy of the neutron).
\end{itemize}
The achieved FWHM energy resolution is 14$\pm$2~keV at 4.78~MeV.\newline
%
%
%

This detector proved to be able to tag neutrons with kinetic energy varying from few meV up to some MeV, assuming that the energy transfer from the neutron to the reaction products is linear. The measured light yield of $^3$H~+~$^4$He interactions in the detector is 0.122$\pm$0.022~keV/MeV at 4.78~MeV, this is estimated considering the amount of light produced by thermal neutron absorption on $^6$Li and the energy released during the reaction. We are not able to disentangle the light signals produced by the two particles, because of the nature of the process (the two particles are produced at the same time), we are just able to give a cumulative LY.\newline

\subsection{Calibration with an $\alpha$ source}
During the testing of our Li$_2$MoO$_4$ crystal, we have also performed a calibration with a $^{147}$Sm $\alpha$ source. The crystal was faced to a smeared Sm $\alpha$ source. The isotope that we were interested in observing was $^{147}$Sm, which $\alpha$-decays with a $Q$-value at 2.31~MeV. The choice of such a low energy $\alpha$ was driven by the fact that we were interested in studying the discrimination power for $\alpha$ and $\beta$/$\gamma$ events in the same energy range, looking at the combination of the thermal and light channels.\newline
\begin{figure}[tb]
\centering
\includegraphics[angle=0, width=0.75\textwidth]{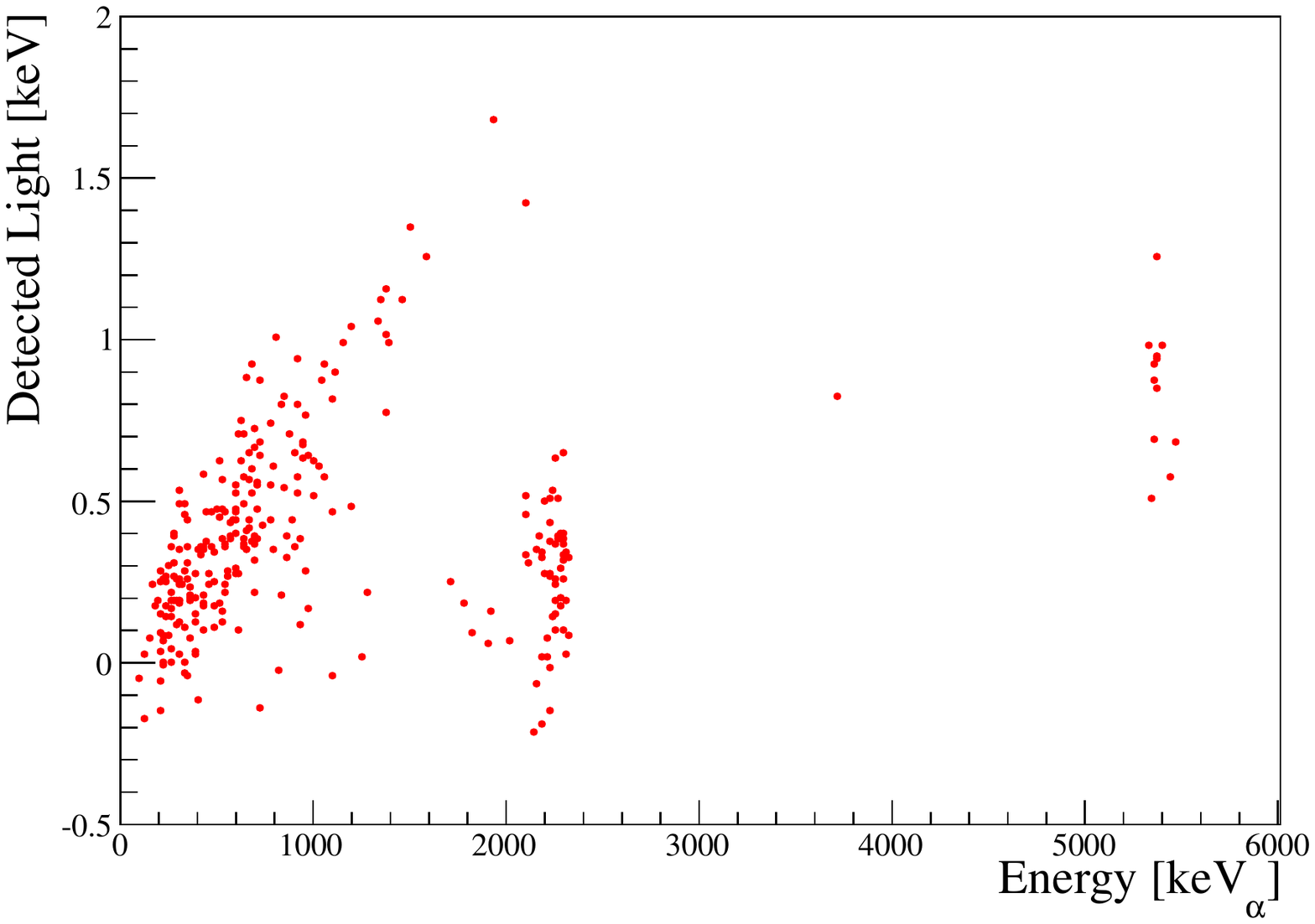}
\caption{Light vs. Heat scatter plot for a 83~h calibration measurement with a Sm $\alpha$ source.}
\label{fig:sm} 
\end{figure}

According to the "Birks law"~\cite{Birks}, the scintillation light produced by $\alpha$ particles is smaller compared to an electron because of its larger d$E$/d$x$. The large density energy deposit of $\alpha$ particles can induce saturation effects in the scintillator. The energy dependence of $\alpha$ particle LY is a consequence of the energy dependence of the stopping power. As previously shown, LY of $\beta$/$\gamma$ events is independent of the energy deposit over a wide energy range (up to 1.46~MeV), we suppose they do not suffer of saturation effects due to their low d$E$/d$x$.\newline

In Fig.~\ref{fig:sm} is shown the calibration measurement. In the band with the lower LY there are two lines, the first one, at 2.25~MeV, is produced by the $\alpha$ source, while the second at higher energy is caused by a $^{210}$Po surface contamination of the crystal at 5.3~MeV.\newline

The computed LY($^{210}$Po) estimated at 5.3~MeV is 0.182$\pm$0.023~keV/MeV, while for the $^{147}$Sm $\alpha$ particle is LY($^{147}$Sm)~=~0.127$\pm$0.013~keV/MeV. These values are estimated for the first time, and they are smaller compared to other important scintillating bolometers used for rare events searches, it is the $\sim$70\% of ZnMoO$_4$~\cite{ZnMoO4}  and the $\sim$5\% of CdWO$_4$~\cite{CdWO4}.\newline

By means of the two $\alpha$ lines shown in the scatter plot, we are able to estimate the LY$_\alpha$ as a function of the energy deposit. Fitting the $\alpha$ band with a second-order polynomial function, forcing that at zero energy there is no light production, we can estimate the LY$_{\alpha}$ for different $\alpha$ energy deposits. We obtain the following equation:
\begin{equation}
LY_{\alpha} (E_{\alpha}) = (0.087\pm0.026) + (0.018\pm 0.007)\cdot E_{\alpha}[MeV] \; \; \; keV/MeV.
\end{equation}
The LY$_{\alpha}$ can not be directly compared to the LY$_{\beta / \gamma}$ evaluated in Sec.~\ref{gamma_calib}, because they have been estimated with two independent measurements with different experimental set-ups, hence different light collection efficiencies.\newline

Before the setting-up of the samarium there was no clear evidence of $^{210}$Pb contamination in the bulk of the crystal (the $\gamma$ calibrations, described in Sec.~\ref{gamma_calib}, did not show any $\alpha$ contaminations during the 87~h measurement). During the installation of the Sm source, the detector was unwillingly contaminated by $^{210}$Pb. Li$_2$MoO$_4$ crystals are hygroscopic, thus when we were assembling our detector, the surface of the absorber was contaminated by $^{210}$Pb nuclei implantation on the surface, probably because the $^{222}$Rn content in the working environment was not minimized, as explained in~\cite{SF}.\newline

From the data acquired during the $\alpha$ calibration, we estimate also the scintillation Quenching Factor for $\alpha$ particles which is defined as:
\begin{equation}
QF_{\alpha}(E)=LY_\alpha(E)/LY_{\beta / \gamma}(E).
\end{equation}
For an $\alpha$ of 5.3~MeV ($^{210}$Po) the scintillation QF$_\alpha$ is 0.42$\pm$0.03, for this estimation we assumed that the LY$_{\beta / \gamma}$ is constant from 1460~keV up to 5.3~MeV. This value is larger compared to other compounds containing Mo: it is 2.5 times larger than that of ZnMoO$_4$~\cite{ZnMoO4} and PbMoO$_4$~\cite{PbMoO4}. For this reason Li$_2$MoO$_4$ seems to be a candidate for DBD searches in $^{100}$Mo, thanks to the good particle discrimination, but obviously highly performing LD are needed.

\section{Internal contaminations}
The Li$_2$MoO$_4$ crystal was grown by means of the Czochralski technique starting from MoO$_3$ and Li$_2$CO$_3$ powders. This crystal was the same used in~\cite{Radio_LiMo}, where the authors have evaluated the intrinsic radiopurity level by means of High Purity Germanium $\gamma$ spectroscopy. In this work, we present improved limits on the main primordial contaminants, $^{238}$U and $^{232}$Th, by few orders of magnitude. The contamination level of the crystal has been evaluated by looking at the $\alpha$ band of the energy scatter plot, where the background level is lower compared to the $\beta$/$\gamma$ one, and the sensitivity is higher.\newline

\begin{figure}[tb]
\centering
\includegraphics[angle=0, width=0.75\textwidth]{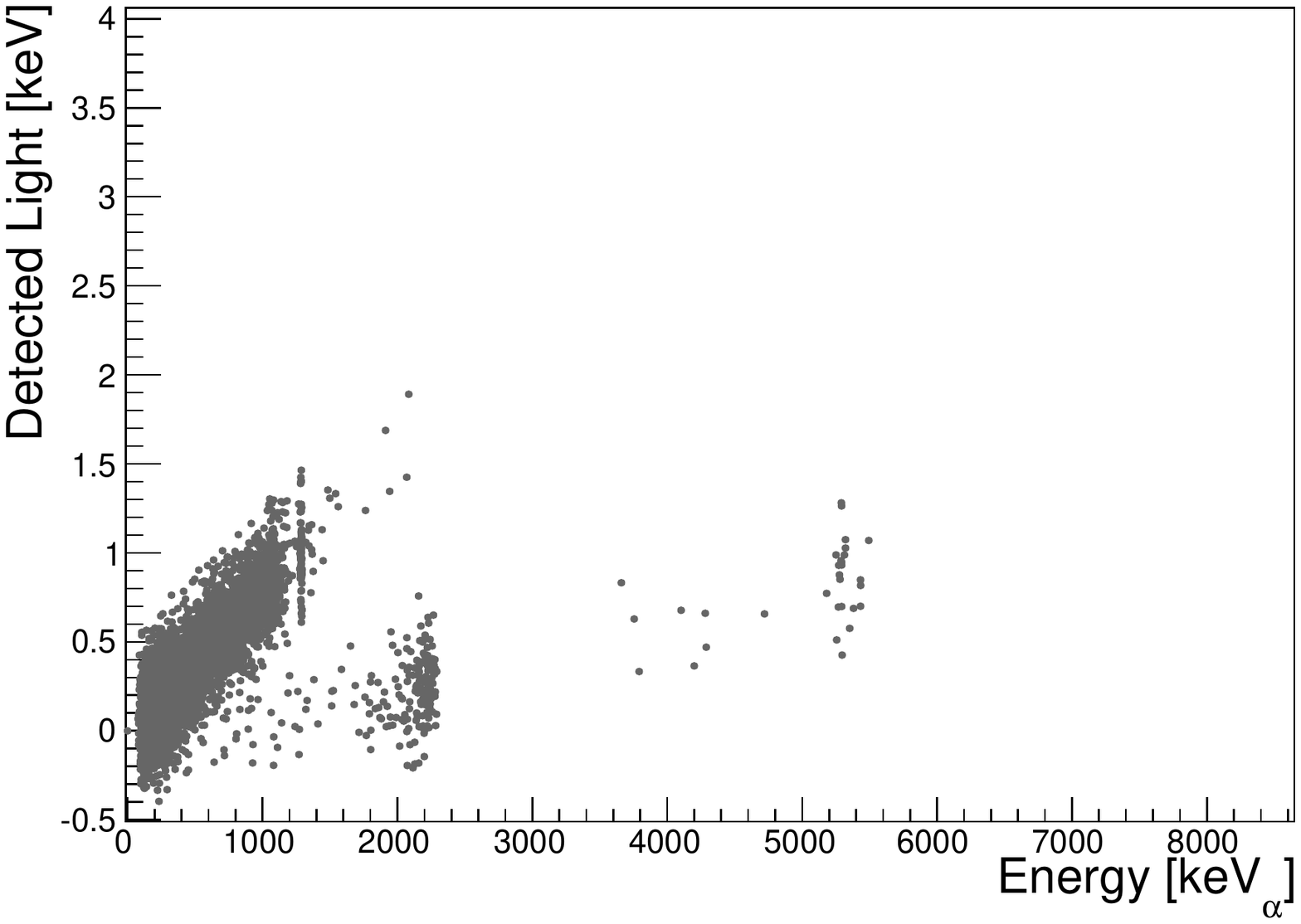}
\caption{Light vs. Heat scatter plot for the 344~h background measurement.}
\label{fig:bkg} 
\end{figure}

The detector was operated in background conditions for 344~h, in Fig.~\ref{fig:bkg} is shown the acquired statistics. During the measurement we were not able to remove the $^{147}$Sm source, nevertheless we were able to evaluate the internal contaminations of the crystal by looking at higher energies of the $\alpha$ energy spectrum (higher than 2.3~MeV). The heat axis of the scatter plot is calibrated with the $\alpha$ lines. As already mentioned in the previous section the two $\alpha$ lines are produced by $^{147}$Sm at 2.25~MeV, expressly deposited on the set-up, and by $^{210}$Po caused by a surface contamination of the crystal.\newline

The $\beta$/$\gamma$ band shows a line at 1.28~MeV, this is produced by $^{40}$K, but it is expected to be found at 1.46~MeV. This miscalibration is not produced by a wrong extrapolation of the calibration function, but by a different mechanism of the energy conversion inside the absorber~\cite{lecoq}. In fact, when a particle deposits energy in the absorber, this can be released through the emission of scintillation photons, the production of thermal phonons or the stimulation of not-detectable channels (e.g. lattice imperfections); the same phenomenon is observed in other crystals~\cite{PbWO4}\cite{CdWO4}. However, the energy calibration of the $\beta$/$\gamma$ scale can be corrected by providing a different calibration for this band.\newline

The detector has a high radiopurity level of internal contaminations, in Table~\ref{tab:contaminations} are shown the results. The secular equilibrium for the $^{232}$Th chain and for the first part of the $^{238}$U chain is assumed. The given limits are more stringent compare to the ones given in~\cite{Radio_LiMo}. We are not able to estimate the internal content of $^{40}$K due to the intense $^{40}$K contamination of the experimental set-up. We would like to stress the fact that the $^{210}$Pb contamination was caused by a not careful handling of the crystal during the assembly of the experimental set-up, it is not an internal background source of the crystal. 
\begin{table}
\caption{Evaluated internal radioactive contaminations for the Li$_2$MoO$_4$ crystal. Limits are computed at 90\% C.L., and are more stringent compared to the given in~\cite{Radio_LiMo}, where the same crystal was studied.} 
\begin{center}
\begin{tabular}{lcc}
\hline\noalign{\smallskip}
Chain & Nuclide  & Activity \\ 
            & & [$\mu$Bq/kg] \\
\noalign{\smallskip}\hline\noalign{\smallskip}
$^{232}$Th & $^{232}$Th & $<$~94 \\
\noalign{\smallskip}\hline\noalign{\smallskip}
$^{238}$U & $^{238}$U & $<$~107 \\
 & $^{210}$Pb & 729$\pm$160 \\
\noalign{\smallskip}\hline
\end{tabular}
\label{tab:contaminations} 
\end{center}
\end{table}

\section{Conclusions}	
We successfully tested a 33~g Li$_2$MoO$_4$ crystal for more than 400~h as a scintillating bolometer. Calibrations with different sources were performed, energy resolution and LY for the different interacting particles were estimated. We demonstrated the detector ability to identify neutrons from $\beta$/$\gamma$ and from $\alpha$ events. This feature is important for direct dark matter searches, where high energy neutrons may induce an unavoidable background in the region of interest.\newline

The high radiopurity level and the scintillation properties make the studied compound a good candidate for the search of double-beta decay in $^{100}$Mo, also because of the large fraction of isotope source with respect to the total mass of the molecule, and for $^7$Li solar axions search.

\acknowledgments
This project was partially carried out in the frame of the LUCIFER project, funded by the European Research Council under the European Union's Seventh Framework Program (FP7/2007- 2013) ERC grant agreement no. 247115. This work was also supported by the ISOTTA project, funded within the ASPERA 2$^{nd}$ Common Call for R\&D Activities.\newline

Thanks are due to the LNGS mechanical workshop and in particular to E. Tatananni, A. Rotilio, A. Corsi, and B. Romualdi for continuous and constructive help in the overall set-up construction. Finally, we are especially grateful to M. Perego and M. Guetti for their invaluable help.


\begin{thebibliography}{99}

\bibitem{Ver12}  J.D. Vergados, H. Ejiri, F. Simkovic, {\it Theory of neutrinoless double-beta decay,
                 Rep. Prog. Phys. {\bf 75} (2012) 106301.}
\bibitem{Giu12}  A. Giuliani, A. Poves, {\it Neutrinoless Double-Beta Decay,
                 Adv. High En. Phys. (2012) 857016.}
\bibitem{Gom12}  J.J. Gomez-Cadenas et al., {\it The search for neutrinoless double beta decay,
                 Riv. Nuovo Cim. {\bf 35} (2012) 29.}
\bibitem{Sim12}  L. Simard et al., {\it The NEMO-3 results after completion of data taking,
                 J. Phys.: Conf. Ser. {\bf 375} (2012) 042011.}
\bibitem{Pir06}  S. Pirro et al., {\it Scintillating Double-Beta-Decay Bolometers,
                 Phys. At. Nucl. {\bf 69} (2006) 2109.}
\bibitem{Lee11}  S.J. Lee et al., {\it The development of a cryogenic detector with CaMoO$_4$ crystals
                 for neutrinoless double beta decay search, Astropart. Phys. {\bf 34} (2011) 732.}
\bibitem{ZnMoO4} J.W. Beeman et al., {\it Performances of a large mass ZnMoO$_4$ scintillating bolometer
                 for a next generation $0\nu$DBD experiment, Eur. Phys. J. C {\bf 72} (2012) 2142.}

\bibitem{PbMoO4}
F.A.~Danevich et al., {\it Feasibility study of PbWO$_4$ and PbMoO$_4$ crystal scintillators for cryogenic rare
events experiments}, {\em Nucl. Instrum. Meth. A} {\bf 622} (2010) 608.                             
                 
                 
\bibitem{Bee12b} J.W. Beeman et al., {\it A next-generation neutrinoless double beta decay experiment
                 based on ZnMoO$_4$ scintillating bolometers, Phys. Lett. B {\bf 710} (2012) 318.}
\bibitem{Bar13}  O. Barinova et al., {\it Czochralski grown large diameter Li$_2$MoO$_4$ single crystals,
                 Poster at 17th Int. Conf. on Crystal Growth and Epitaxy ICCGE-17, Warsaw, Poland,
                 11-16 August 2013.}
\bibitem{DM_review}  G. Bertone, D. Hooper, J. Silk, {\it Particle dark matter: Evidence, candidates and
                 constraints, Phys. Rept. {\bf 405} (2005) 279.}
\bibitem{Fen10}  J.L. Feng, {\it Dark Matter Candidates from Particle Physics and Methods of Detection,
                 Ann. Rev. Astron. Astrophys. {\bf 48} (2010) 495.}
\bibitem{Kim10}  J.E. Kim, G. Carosi, {\it Axions and the strong CP problem,
                 Rev. Mod. Phys. {\bf 82} (2010) 557.}
\bibitem{Resonant_axions}  S. Moriyama, {\it Proposal to Search for a Monochromatic Component of Solar
                 Axions Using $^{57}$Fe, Phys. Rev. Lett. {\bf 75} (1995) 3222.}
\bibitem{Li7_axions}  M. Krcmar et al., {\it Search for solar axions using $^7$Li,
                 Phys. Rev. D {\bf 64} (2001) 115016.}
\bibitem{Bel12}  P. Belli et al., {\it Search for $^7$Li solar axions using resonant absorption
                 in LiF crystal: Final results, Phys. Lett. B {\bf 711} (2012) 41.}
\bibitem{Radio_LiMo}  O.P. Barinova et al., {\it Intrinsic radiopurity of a Li$_2$MoO$_4$ crystal,
                 Nucl. Instrum. Meth. A {\bf 607} (2009) 573.}
\bibitem{First_LiMo}  O.P. Barinova et al., {\it First test of Li$_2$MoO$_4$ crystal as a cryogenic
                 scintillating bolometer}, Nucl. Instrum. Meth. A {\bf 613} (2010) 54.
                 

\bibitem{LD}
 J.~Beeman et~al., \emph{ Characterization of bolometric Light Detectors for rare event searches}, submitted to {\em J. Inst.} [arXiv:1304.6289].

\bibitem{QINO}
 E.~Andreotti et~al., \emph{ $^{130}$Te neutrinoless double-beta decay with CUORICINO}, {\em Astropart. Phys.} {\bf 34} (2011) 822.

\bibitem{Gatti}
E.~Gatti and P.~F. Manfredi, {\it Processing the signals from solid state detectors in elementary particle physics},  {\em Riv. Nuovo Cimento} {\bf 9} (1986) 1.

\bibitem{Radeka}
V.~Radeka and N.~Karlovac, {\it Least-square-error amplitude measurement of pulse signals in presence of noise},  {\em Nucl. Instrum. Methods} {\bf 52} (1967) 86.

\bibitem{light_sync}
G.~Piperno, S.~Pirro, M.~Vignati, {\it Optimizing the energy threshold of light detectors coupled to luminescent bolometers}, {\em J. Inst.} {\bf 6} (2011) P10005.

\bibitem{Lithium-6}
J.~Gironnet et~al., {\it Neutron spectroscopy $^6$LiF bolometers}, {\em AIP Conf. Proc.} {\bf 1185} (2009) 751.

\bibitem{PbWO4}
 J.~Beeman et~al., {\it New experimental limits on the $\alpha$ decays of lead isotopes}, {\em Eur. Phys. J. A} {\bf 49} (2013) 50.

\bibitem{XS_Li6}
Japan Atomic Energy Agency, {\it Nuclear Data Center}, http://wwwndc.jaea.go.jp/.

\bibitem{Birks}
J.B.~Birks, \emph{The Theory and Practice of Scintillation Counting}, Pergamon, London, 1964.

\bibitem{CdWO4}
C.~Arnaboldi et~al., {\it CdWO$_4$ scintillating bolometer for Double Beta Decay: Light and Heat anticorrelation, light yield and quenching factors}, {\em Astropart. Phys.} {\bf 34} (2010) 143.

\bibitem{SF}
M.~Clemenza, C.~Maiano, L.~Pattavina, and E.~Previtali, {\it {Radon-induced
  surface contaminations in low background experiments}},  {\em Eur. Phys. J. C}  {\bf 71} (2011) 1805.


\bibitem{lecoq}
P.~Lecoq et~al., {\it Inorganic Scintillators for Detector Systems: Physical Principles and Crystal Engineering}, Springer, Berlin, 2006.



\bibitem{CaMoO4}
C.~Arnaboldi et~al., {\it A novel technique of particle identification with bolometric detectors}, {\em Astropart. Phys.} {\bf 34} (2011) 797.


 
\end{thebibliography}
\end{document}